\documentclass[a4paper]{jpconf}
\usepackage{graphicx}
\def\beq{\begin{equation}}
\def\eneq{\end{equation}}
\def\bea{\begin{eqnarray}}
\def\enea{\end{eqnarray}}
\begin{document}
\title{Ho\v rava--Lifshitz gravity with detailed balance}

\author{Daniele Vernieri and Thomas P. Sotiriou}

\address{SISSA, Via Bonomea 265, 34136, Trieste, Italy {\rm and} INFN, Sezione di Trieste, Italy}

\ead{daniele.vernieri@sissa.it, sotiriou@sissa.it}

\begin{abstract}
Ho\v rava--Lifshitz gravity with ``detailed balance'' but without the projectability assumption is discussed. It is shown that detailed balance is quite efficient in limiting the proliferation of couplings in Ho\v rava--Lifshitz gravity, and that its implementation without the projectability assumption leads to a theory with sensible dynamics. However, the (bare) cosmological constant is restricted to be large and negative.
\end{abstract}

\section{Introduction}

Ho\v rava--Lifshitz gravity has been proposed as a power-counting renormalizable gravity theory, and a potentially ultra-violet (UV) complete theory of quantum gravity  \cite{arXiv:0901.3775}. The improved UV behaviour is achieved by giving up lorentz symmetry and adding terms to the action that are higher order in spatial derivatives, which suitably modifies the propagators. The theory is most straightforwardly defined in a space-time decomposition (ADM splitting) where a preferred foliation has been imposed and $N$ is the lapse, $N_i$ is the shift and $g_{ij}$ the 3-metric induced on the spacelike hypersurfaces. The action of the theory is then
\beq
S=S_K-S_V\,,
\eneq
where the kinetic term $S_K$ is given by
\beq
S_k=\frac{2}{k^2}\int dtd^3x\sqrt{g} N K_{ij}G^{ijkl}K_{kl},
\eneq 
$K_{ij}$ is the extrinsic curvature of the spacelike hypersurfaces, $k$ a coupling constant with suitable dimensions and 
\beq
G^{ijkl}=\frac{1}{2}\left(g^{ik}g^{jl}+g^{il}g^{jk}\right)-\lambda g^{ij}g^{kl}\,,
\eneq
is the generalized DeWitt metric.
The potential term can be written as
\beq
S_V=\frac{k^2}{8}\int dtd^3x\sqrt{g} N \, V[g_{ij},N].
\eneq
The action is required to be invariant under the subclass of diffeomorphisms that leave the foliation intact, {\em i.e.}~$t\to \tilde{t}(t)$ and $x^i\to \tilde{x}^i(t,x^i)$, so in principle $V$ should include all terms compatible with this symmetry. In addition, $V$ should contain terms which are at least 6th order in spatial derivatives for the theory to have the desirable UV properties  \cite{arXiv:0901.3775,arXiv:0902.0590,arXiv:0912.4757}. 

An unappealing feature of the full theory is that, without any further restrictions or symmetries, $V$ will contain a very large number of terms (and independent couplings). Two restrictions have been considered as the means to limit the proliferation of couplings in the original proposal: {\it projectability} and {\it detailed balance}. Projectability sums up to the requirement that $N=N(t)$, whereas detailed balance requires $V$ to be derived from a ``superpotential" (see below). 

An important characteristic of the theory is that it propagates not only a spin-2, but also a spin-0 mode.
The assumption of projectability, with or without detailed balance \cite{arXiv:0904.4464} leads to pathologies in the dynamics of the scalar mode, in particular instabilities and strong coupling at low energies \cite{arXiv:0905.2798,Charmousis:2009tc,Blas:2009yd,Wang:2009yz,Afshordi:2009tt,Koyama:2009hc}. For this reason, here we discuss a version of the theory which satisfies the detailed balance condition without assuming projectability.

This proceedings contribution is based on Ref.~\cite{Vernieri:2011aa}, where a more detailed discussion can be found. See also Ref.~\cite{arXiv:1010.3218} for a brief review on the various version of Ho\v rava--Lifshitz gravity.

\section{Detailed balance without projectability}

Detailed balance prescribes that the potential $V$ is derived from a superpotential $W$ as follows
\beq
V=\frac{1}{\sqrt{g}}\frac{\delta W}{\delta g_{ij}}G_{ijkl}\frac{1}{\sqrt{g}}\frac{\delta W}{\delta g_{kl}}\,, \label{DB}
\eneq
where $G_{ijkl}$ is the inverse of the DeWitt metric $G^{ijkl}$.
Once $W$ has been chosen the action is fully determined. $W$ can be a functional of the metric $g_{ij}$ but also of the lapse $N$, and in particular of the combination $a_i=\partial_i\mbox{ln}N$, if invariance under foliation preserving diffeomorphisms is imposed \cite{arXiv:0909.3525}. Furthermore, in order to have a power-counting renormalizable theory one needs at least sixth order spatial derivatives in the action \cite{arXiv:0901.3775,arXiv:0902.0590,arXiv:0912.4757}, thus the superpotential $W$ must contain at least  third order spatial derivatives. Then, the most general $W$ containing all of the possible terms up to third order spatial derivatives is
\beq
W=\frac{M_{\rm pl}^2}{2M_6^2}\int{\omega_3(\Gamma)}+\frac{M_{\rm pl}^2}{M_4}\int{d^3x\sqrt{g}\left[R-2\xi(1-3\lambda)M_4^2\right]}+\beta\int{d^3x\sqrt{g}\,a_i a^i}, \label{W}
\eneq 
where $\omega_3(\Gamma)$ is the gravitational Chern-Simons term, $M_{\rm pl}$, $M_6$, $M_4$ and $\beta$ have dimensions of a mass, whereas $\xi$ is a dimensionless coupling. 
The corresponding action is
\bea
\label{faction}
S_H&=&\frac{M_{\rm pl}^2}{2}\int dt d^3x \sqrt{g}N\Bigg\{ K_{ij}K^{ij}-\lambda K^2+\xi R-2\Lambda+\eta \,a^ia_i-\frac{1}{M_4^2}R_{ij}R^{ij}+\frac{1-4\lambda}{4 (1-3\lambda)} \frac{1}{M_4^2}R^2\nonumber \\
&&+\frac{2\eta}{\xi M_4^2}\left[ \frac{1-4\lambda}{4 (1-3\lambda)} R a^ia_i-R_{ij}a^i a^j \right]-\frac{\eta^2}{4\xi^2 M_4^2}\frac{3-8\lambda}{1-3\lambda} (a^ia_i)^2+\frac{2}{M_6^2 M_4}\epsilon^{ijk}R_{il}\nabla_j R^l_k \nonumber\\
&&+\frac{2\eta}{\xi M_6^2 M_4}C^{ij}a_ia_j-\frac{1}{M_6^4}C_{ij}C^{ij}\Bigg\}\,, 
\enea
where $C_{ij}$ is the Cotton--York tensor and 
\beq
\eta=\frac{\beta\,\xi M_4}{M_{\rm pl}^2}\,, \qquad \Lambda=\frac{3}{2}\xi^2(1-3\lambda)M_4^2. \label{CC}
\eneq
For $\xi,\lambda\sim1$, which are the values these parameters have in general relativity, the bare cosmological constant, $\Lambda$, is negative \cite{arXiv:0904.4464}. Moreover, the magnitude of $\Lambda$ is directly related to $M_4$ \cite{arXiv:0907.3121,Vernieri:2011aa} which is the mass scale at which Lorentz-violating effects become manifest as modifications to the dispersion relation. The mildest constraint on the value of $M_4$, which comes by the fact that the gravitation interaction has been tested down to mm scales,  is $M_4\geq 1\div 10{\rm meV}$.\footnote{Stronger constraints can be obtained for $M_4$ but depend on the details of how Lorentz violations percolate into the matter sector, see {\em e.g.}~Ref.~\cite{Liberati:2012jf}.} Thus, the bare cosmological constant is at best 60 orders of magnitude bigger than the observed value.  

The overall value of the cosmological constant would be the sum of the bare cosmological constant and the contribution from the vacuum energy of matter fields. Therefore, the size and not the sign of the bare cosmological constant is the main concern here. The contribution of the vacuum energy is not known with any certainty, but one can only hope for a miraculous cancelation if the overall value is to match the observed value. There is no known reason to expect such a cancelation (see, however, Ref.~\cite{arXiv:0907.3121}).

\section{Perturbations and power-counting renormalizability}

In order to highlight that the magnitude of the cosmological constant is the only real concern and that the action constructed here describes a dynamically sensible theory, we will proceed as follows: we will set $\Lambda=0$ by fiat (to simplify the analysis) and consider perturbations around a flat background. We focus in particular to scalar perturbations, as the main concern is the dynamics of the scalar mode. Thus we have
\beq
N=1+\alpha\,, \qquad    N_i=\partial_i y\,, \qquad g_{ij}=e^{2\zeta}\delta_{ij}\,,
\eneq
and at quadratic order the action becomes
\beq  \label{quad}
S^{(2)}\!\!=\!\!\frac{M_{\rm pl}^2}{2}\int dt d^3x \bigg\{\frac{2(1-3\lambda)}{1-\lambda}\dot{\zeta}^2+2\xi\left(\frac{2\xi}{\eta}-1\right)\zeta\partial^2\zeta -\frac{2(1-\lambda)}{1-3\lambda}\frac{1}{M_4^2}(\partial^2\zeta)^2\bigg\}\,,
\eneq
where we have integrating out the nondynamical degrees of freedom $\alpha$ and $y$.
The corresponding dispersion relation for the scalar is
\beq
\omega^2=\xi \left(\frac{2\xi}{\eta}-1\right)\frac{1-\lambda}{1-3\lambda}p^2+\frac{1}{M_4^2}\left(\frac{1-\lambda}{1-3\lambda}\right)^2p^4. \label{disp}
\eneq
Then we can conclude that there is a suitable choice of the parameters such that the scalar has positive energy (the right sign is determined by the sign of the kinetic term of the spin-2 graviton), and at the same time is classically stable:
\beq
2\xi>\eta>0\,.
\eneq
The coefficient of the $p^4$ term in eq. (\ref{disp}) is manifestly positive, so there are no short-wavelength instabilities. However, the dispersion relation does not contain any sixth order terms. This is because the only sixth order term in action (\ref{faction}) does not contribute to the dynamics of the scalar more. The absence of sixth order terms in the dispersion relation poses a threat for renormalizability. However, a straightforward resolution is to add to the superpotential $W$ fourth order terms, so as to generate sixth (and eight order terms) in the action, which would restore the appealing renormalizability properties of the theory. The minimal consistent prescription would be to add to $W$ all possible fourth order terms that respect invariance under foliation preserving diffeomorphisms, but also impose parity invariance. Then the terms added would be 
\beq
R^2\,, \quad R^{\mu\nu}R_{\mu\nu}\,, \quad R \nabla^i a_i\,, \quad R^{ij}a_i a_j\,, \quad R a_i a^i\,,\quad (a_i a^i)^2\,, \quad (\nabla^i a_i)^2\,, \quad a_i a_j \nabla^i a^j\,.
\eneq
One would be left with 12 free couplings, which is an appreciable improvement with respect to the theory without detailed balance.

\section{Conclusions}

We have put together an action for Ho\v rava--Lifshitz gravity that satisfies detailed balance without assuming projectability. This version appears to lead to sensible dynamics for the scalar mode and be free of other pathologies that plague the projectable version of Ho\v rava--Lifshitz gravity. Additionally, it has the appealing renormalizability properties of the most general non-projectable theory, with about an order of magnitude less independent couplings.

However, imposing detailed balance leads to a large and negative bare cosmological constant. Barring a miraculous cancelation with the (yet to be understood) vacuum energy contribution to the cosmological constant, this is obviously hard to reconcile with low energy gravitational phenomenology. An additional caveat in the use of detailed balance is the fact that it is not clear if or why it should be robust against radiative corrections. The implementation of detailed balance presented here leads to a consistent theory which is free of pathologies and, therefore, allows to consider this question rigorously.

\end{document}